\documentstyle{mn}
\newcommand{\be}{\begin{equation}}
\newcommand{\ee}{\end{equation}}

\title{Stellar Forensics: I - Cooling Curves}
\author[B. M. S. Hansen \& E. S. Phinney]
{Brad M. S. Hansen$^{1,2}$\thanks{email:{\bf hansen@cita.utoronto.ca}} \& E. Sterl Phinney$^2$
\\
$^1$ Canadian Institute for Theoretical Astrophysics, University of
Toronto, Toronto, ON M5S 3H8, Canada \\
$^2$ Theoretical Astrophysics, 130-33, California Institute of Technology,
Pasadena, CA 91125}
\date{28 August 1997}

\pagerange{\pageref{firstpage}--\pageref{lastpage}}
\pubyear{1997}
\begin{document}

\maketitle
\label{firstpage}

\begin{abstract}
The presence of low mass, degenerate secondaries in millisecond pulsar binaries offers
the opportunity to determine an age for the binary system independent of the rotational
properties of the pulsar. To this end, we present here a detailed calculation of the evolution of
a grid of low mass ($< 0.5 \, \rm M_{\odot}$) helium core white dwarfs. We investigate the
effects of different Hydrogen layer masses and provide results for well-known optical
band-passes. We supplement the OPAL opacity calculations with our own calculations for
low effective temperatures ($ T_{\rm eff} < 6000 \, K$) and also provide fitting formulae for
the gravity as a function of mass and effective temperature. In paper~II we shall apply these
results to individual cases.
\end{abstract}

\begin{keywords}
binaries:general --- stars: evolution --- stars: fundamental parameters --- white dwarfs
\end{keywords}

\section{Introduction}

Studies of binary pulsars  (see Phinney \& Kulkarni 1994 for references)
 and close double degenerate systems (Marsh, Dhillon \& Duck 1995) have discovered a number
of
 low mass objects. The masses of these optically faint stars are estimated to be
$\sim$ 0.1 - 0.5 $\rm M_{\odot}$. Thus, it is believed that these are helium
core white dwarfs, which are not massive enough to ignite core helium burning
and burn to carbon (Sweigart \& Gross 1978; Mazzitelli 1989). Such stars arise
in binaries because the progenitor is disrupted by Roche lobe overflow
at some point in its natural
evolution, losing its envelope and leaving behind a low-mass, degenerate
helium star (Kippenhahn, Kohl \& Weigart 1967).

Many millisecond pulsar binaries contain such dwarfs and, as such, their cooling
ages offer an estimate of the age of the system, independent of the pulsar
spin-down age and thus an interesting check of the traditional assumptions made about
the ages of millisecond pulsars.

Our aim in this paper is to present an extensive grid of cooling models covering
the parameter space of low-mass helium core white dwarfs, much in the spirit of
previous studies of the more ubiquitous carbon/oxygen core white dwarfs 
(
Wood~1992; D'Antona \& Mazzitelli 1990 and references therein). In paper~II we 
apply these models to the observations of the low-mass binary pulsar population.



In section~\ref{Cooling} we briefly review the physical mechanisms which contribute to
the white dwarf cooling process and discuss their relative importance. Section~\ref{LCS}
describes our calculations of the low-temperature opacities necessary to obtain accurate
cooling sequences for the oldest systems.
 In section~\ref{Code} we describe our numerical model and the tests
of the code against other models from the literature. Finally, in section~\ref{ResultsI}
we present our cooling sequences and describe the details of the cooling models.

\section{White Dwarf Cooling}

\label{Cooling}

The basic qualitative picture of white dwarf cooling is a well-known one, going back to
 Mestel (1952). The star is supported by the pressure of a degenerate electron gas while
the heat content is dominated by the thermal reservoir of
 non-degenerate ions, which can form  a 
gas or a crystalline solid, depending on the Coulomb coupling parameter $ \Gamma = (Z
e)^2/a k T$, where $a$ is the
radius of the Wigner-Seitz sphere surrounding each ion. While the luminosity is dominated by
the loss of thermal energy, young, hot white dwarfs may also have a contribution due to residual
hydrogen burning at the base of the hydrogen envelope.

The energy transport throughout most of the star is dominated by conduction due to the
degenerate electrons, which keeps the core almost isothermal.
The core is surrounded by a thin, non-degenerate envelope where the energy transport is
by radiative diffusion and, at late times, by convection (B\"{o}hm et al. 1977).
 This is the region of least efficient energy transport and thus it determines
 the rate at which the star cools. Hence, the chemical
composition of these outer layers is important. The high gravities of white dwarfs leads to
gravitational settling (Schatzmann 1958; Dupuis et al. 1992 and references therein)
 which results in chemical separation in the
white dwarf envelope and hence the atmospheric opacity is dominated by the lightest species
remaining on the surface, usually either hydrogen or helium. 

Since the cooling is determined by the radiative transport, the accuracy of the cooling curves
will be dependant on the accuracy of the opacities we use. Over most of the cooling sequence
we use the radiative opacities from the OPAL group (Rogers \& Iglesias 1992), which are good down
to $ T = 6000$~K. For conductive opacities we use the results of Itoh et al. (1983) and
 Mitake, Ichimaru \& Itoh (1984)
 for the
region where the ions are gas or liquid and the electrons degenerate (helium cores don't reach
 crystallization temperatures within a Hubble time). At temperatures and densities
not covered by the Itoh results, we use the older conductivities of  Hubbard and Lampe (1969).
This still leaves the region $ T<6000 \, \rm K$ unaccounted for. It is important because many of the
objects we will discuss in paper~II have $ T_{\rm eff} \sim 4000 \, \rm K$. 
To address this issue, we have
calculated $ Z=0$ opacities for an arbitrary mix of hydrogen and helium using primarily the 
input physics of Lenzuni, Chernoff and Salpeter (1991), henceforth called LCS. In particular, we
calculate the opacities from the collisionally induced absorption
(CIA) by molecular hydrogen.
We describe this briefly below.

\section{Low Temperature $Z=0$ Opacities}
\label{LCS}

We calculate the opacity of a gas containing H, He, H$_2$, H$^-$,He$^-$,H$^+$,H$_2^+$,H$_3^+$,He$^+$ and
electrons. We assume ideal gas and LTE. From the equation of state of  Saumon,
Chabrier and Van Horn (1995),
 we determine that the assumption of an ideal gas is good for
$ \rho<10^{-2} \rm g \, cm^{-3}$. The partition functions and cross-sections are taken largely from
LCS. For the calculation of collisionally induced  molecular opacities we used the fits of LCS for the
 H$_2$-H$_2$ 
roto-vibrational transitions (with corrections for typographical errors for which we thank
Dr. Chernoff), after checking it with the original code  (Borysow and Frommhold 1990),
kindly supplied to us by Dr. A. Borysow. The roto-translational contribution was recalculated using
another, more recent, code (Zhang \& Borysow 1995) again supplied by Dr. Borysow. We also recalculated the 
H$_2$-He opacity following LCS although most of our atmospheres end up with pure H or He compositions.

When we use the composition $X$=0.72 and $Y$=0.28, we find good agreement with LCS except at the
high temperature, high density end $ T \sim 6000-7000$ K and $ \rho \sim 10^{-3} \rm g\, cm^{-3}$, where we have a 10-20\% error. We attribute this to
our LTE treatment of the H$^-$ ion. LCS note that the equilibrium abundance of this
ion is affected by the radiation field at almost any temperature because of its low dissociation energy.
Since the OPAL opacities reach down to $6000$ K and our 
atmospheres are convective in this region  anyway (see Figure~\ref{3Cool2}),
 we opt for simplicity and use our LTE results.

For the case of a pure hydrogen atmosphere of moderate density $ \rho \sim 10^{-5} \rm g \, cm^{-3}$,
 the opacity is dominated above $\sim 3000$ K by the $\rm H^-$ ion and below by the collisionally
induced opacity of H$_2$. As noted by Bergeron, Saumon \& Wesemael (1995), the Rosseland opacity
 goes through a minimum near this temperature. For a pure helium atmosphere He$^-$ and Rayleigh
scattering provide most of what little opacity there is. 

We extend our calculations up to densities $ \rho \sim 1 \, \rm g \, cm^{-3}$.
At higher densities, in lieu of an accurate calculation,
 we introduce abrupt
pressure ionization and the opacity is determined by electron conduction at
higher densities.
For the case of a hydrogen atmosphere, this is unimportant because the atmosphere is sufficiently
opaque that convection will dominate by the time these densities are reached. However, for the
case of a helium atmosphere, this is the major source of uncertainty in our cooling times because
the atmosphere is very optically thin and where we place the pressure ionization will determine
the location of the photosphere.

Figure~\ref{hydrogen} shows the contours of constant opacity in our density-temperature
parameter space for pure hydrogen. It also shows the regions where different opacity
tables were used. In Figure~\ref{match}, as an example, we show the opacity as a function of temperature at $\rho = \rm 
10^{-9} g \, cm^{-3}$ for a pure hydrogen atmosphere. There is excellent agreement between our
calculation and 
the OPAL tables in the overlap region.


\section{The Cooling Code}
\label{Code}

We have written a numerical cooling code using the Henyey method (Henyey et al. 1959,1964)
to solve the stellar structure equations. Our outer boundary conditions (obtained using a
grey atmosphere calculation) are implemented using
the Kippenhahn, Weigart \& Hoffmeister (1967) method of triangles. 

Apart from the opacity tables described above, we have used the hydrogen and helium equation
of state of Saumon et al. (1995), supplemented by a Thomas-Fermi model where necessary (at high
densities). For the carbon/oxygen sequences used in our comparisons (see section~\ref{tests}), we also
used the carbon equation of state table of Fontaine, Graboske \& Van Horn (1977). For completeness, we also
include the effects of neutrino losses using the emissivities of Itoh \& Kohyama (1983),
 Itoh et
al (1984) and Munakata, Kohyama \& Itoh (1985). To account for the residual nuclear burning, we incorporate
the cross-sections of  Fowler, Caughlan and Zimmerman (1975). We also use the mixing length
theory of convection (we tested different parameterisations, ML1, ML2 and ML3 (see
 Bergeron, Wesemael \& Fontaine 1992 and references therein), and found no
difference in their effect on the cooling). Appendix~\ref{NumDet} briefly
describes the numerical details of the calculation.

\subsection{Code Tests}
\label{tests}

The dearth of reliable helium white dwarf cooling sequences means that, in order to test our
numerical method properly, we need to include carbon cores so as to test our models against those in the
accepted literature.
We have made comparisons with three well known codes,
adjusting our input physics to approximate the original input physics as closely as possible.
The wide variation in input physics and parameters makes this a non-trivial proposition 
(see
Winget \& Van Horn 1987).  Nevertheless
we obtain satisfactory fits that give us confidence in the accuracy of our numerical scheme and
also
incidentally reinforces the assertion of Winget and Van Horn that the wide variation in published
white
dwarf cooling models is a result of input physics and not numerical treatment. 
 The comparison is shown
in Figure~\ref{compare}. The three models are from  D'Antona \& Mazzitelli (1989),
 Koester \&
Sch\"{o}nberner
(1986) and  Wood (1992). Of particular interest is the comparison with the Wood cooling sequence
because
that is the most up-to-date, using $Z=0$ OPAL opacities.
 We don't expect agreement
 at late times
because the calculations of the opacities for $ T<6000 \rm \, K$ will vary, but the agreement
 for ages less than
a few billion years is excellent. In particular, the arrows indicate the point at which
crystallisation begins, and we do not expect agreement at later times than this because our
treatment of crystallisation is not sufficiently detailed for these purposes (Helium cores
do not crystallise in a Hubble time and thus it is unimportant for this analysis). An updated
version of these curves will presented in a future publication.

The one useful test of our code for the case of a helium white dwarf is a comparison with the
cooling sequence of Iben \& Tutukov (1986), shown in Figure~\ref{it}.
 Their model was for a $0.3 \, \rm M_{\odot}$ star which,
after undergoing two hydrogen shell flashes, has a shell of $2.5 \times 10^{-4} \, \rm M_{\odot}$ of
pure hydrogen during the white dwarf phase. They caution that their opacities are uncertain
at low luminosities, but it provides a useful comparison at least for $\log L/L_{\odot} > -3$.
Their model also demonstrates another uncertainty for our higher luminosity models. The
white dwarf stage really only begins after the end of the second hydrogen shell flash which 
means that our $t=0$ corresponds to $t=10^8$ years in this case. The effect of this uncertainty 
in the starting point for the model is important for $\log  L/L_{\odot} > -3$.

A second uncertainty in the initial conditions is the value of the core temperature at the
beginning of the evolution. Tests of our code show that an uncertainty of $3 \times 10^7$ K
corresponds to an age uncertainty $\sim 10^8$ years, although the exact numbers are model
dependant. We solve this problem by using the results of evolutionary calculations
(Soberman, pers. comm.) of 
the progenitors at a variety of binary separations around a neutron star, using the stellar
evolution code of Eggleton (1971) (see also Pols, Tout \& Eggleton 1995). This provides us with an
estimate of the initial central temperature for a given remnant mass, which we use as a 
starting point for our calculations. For most of the systems that we
address in paper~II, with ages $>$ 1 Gyr,
this uncertainty is unimportant.

There are also uncertainties in the evolution associated with the uncertainties in the
input physics. We investigated the effect of extrapolating the Itoh opacities to lower
densities instead of using the old Hubbard and Lampe opacities for those regions. This
was found to have little impact on the cooling sequences, because the conductivity is
most important in the highest density regions, which is covered by the Itoh tables. A
second uncertainty is the effect of extrapolating the OPAL opacities to higher densities in 
the region $ T \sim 10^4-10^5$ K and $ \rho \sim 10^{-3} - 1 \rm \, g \, cm^{-3}$. This is
not terribly important because the atmosphere is convective in this region, and so the
extrapolation will affect only the depth at which convection terminates.

A major uncertainty of past calculations was for the low temperature ($T <$ 6000 K)
opacities. Our calculations in section~\ref{LCS} are designed to solve this problem, at
least for the hydrogen atmospheres. The problem persists with the helium atmospheres, where
the photospheric opacity is determined by the pressure ionization of helium, in a regime
where accurate
opacity calculations do not exist. For some models, we also have convective mixing of the
elements. We assume that the convective zone is mixed to a uniform composition determined
by the relative mass fractions of hydrogen and helium in the convective zone. When the 
convective zone retreats from the lower layers, we assume that the separation is
instantaneous, i.e., at all times, the composition in the convective zone is determined
by the depth of the convective base (i.e., the mass of helium in the convection zone
relative to the hydrogen envelope mass). The admixture of helium into the hydrogen
atmosphere does not restrict the validity of our opacity calculations at low temperatures.
Because of the extremely low neutral helium opacities, 5\% of hydrogen by mass is still
sufficient to provide enough opacity so that the photospheric pressure lies
well below the pressure ionization value.

\section{Results}
\label{ResultsI}

In this section, we will describe in detail the cooling of a 0.3 M$_{\odot}$ helium core star with
a thick ($ 3 \times 10^{-4} \rm M_{\odot}$) hydrogen envelope, shown
in Figures~\ref{3Cool3} to \ref{3Cool1}. We start our models with an age of
$10^8$ years (the time taken for the Iben \& Tutukov models to reach the end of the last shell flash).
We find that, even for these hydrogen masses and at early ages, the residual nuclear burning contribution
to the luminosity is never more than $\sim 1 \%$ (see Figure~\ref{3Cool3}).
The neutrino luminosity is never more than $\sim 10 \%$
of the photon luminosity. Thus, over the first few $\times 10^6$ years the star completes the contraction
to the white dwarf configuration that it began when the last shell flash ended (see, e.g., Iben \&
Tutukov 1986, Table 1). In these early stages, 
the helium core is only mildly degenerate (central degeneracies $\sim 30$) with $\eta \sim 1$ at the
hydrogen-helium boundary, although the core is already approaching an isothermal state (0.99 of the
stellar mass has $T > 0.5 \, T_{\rm c}$).
The hydrogen envelope is entirely radiative at this stage, so that this is the time at which the star
most resembles the Mestel ideal (see Figure~\ref{3Cool3}). 

When $ T_{\rm eff} \sim 15 000$~K ($\sim 3 \times 10^8$ years), a small convection zone appears
near the surface. This zone remains relatively thin until $ T_{\rm eff} \sim 10^4 $K ( $\sim 10^9$ years),
when it starts to deepen as a consequence of the movement of the hydrogen ionization zone to
greater depths (Figure~\ref{3Cool1}). The recombination of hydrogen also leads to an increase in the photospheric
density (since neutral hydrogen has a smaller opacity than ionized hydrogen). This increase is 
only halted once molecular hydrogen begins to form ($ T_{\rm eff} \sim 5000 $~K, $t \sim 4 \times 10^9$
years, see Figure~\ref{3Cool2}). These photospheric changes also lead to the flattening of the cooling
curves and the increase in the gravity seen in Figures~\ref{Lt} and \ref{gfit}, similar to effects 
in C/O models as described in D'Antona \& Mazzitelli (1990).
 The deepening of the convective zone continues until the base reaches a depth where the
conduction due to degenerate electrons is more efficient than convection.
 This occurs after $3 \times 10^9$ years. At
its deepest extent, the convection zone contains $\sim 5 \times 10^{-5} \rm M_{\odot}$. This is only
a couple of scale heights above the hydrogen/helium interface, so we might expect some small amount
of photospheric
helium contamination (since the convection extends to optical depths $\tau \sim 0.3$) below
$T_{\rm eff} \sim 6000$ K (see also B\"{o}hm et al 1977).
 However, the helium will be present only in trace amounts and will not
affect the cooling.
As the star cools further, it will revert again to a pure hydrogen atmosphere because of the
formation of a radiative buffer zone near the surface below $ T_{\rm eff} \sim 3500$ K. This occurs
because there is a minimum in the atmospheric opacity as the primary opacity contribution changes
from CIA of $\rm H_2$ to $\rm H^-$ absorption. If this minimum value is
low enough, the region of the minimum will be radiative, but
convection will continue to operate both
above and below. This would allow the surface helium to diffuse out of the top convection zone,
leaving the atmosphere once again in a pure hydrogen state.

The features of the above sequence are common to most of the models we discuss here. The primary
differences occur at the lower mass end of the models ($M \sim$ 0.15 M$_{\odot}$).
 The lower masses lead to lower central densities and lower degeneracy, so that the model radii
 are somewhat larger (hence with lower gravities) and convection extends deeper into the cooler
models (because conduction is less efficient). Thus, even for the thickest hydrogen envelopes
considered here\footnote{We have used the upper limit derived by Iben \& Tutukov (1986),
although it is likely to increase somewhat for smaller core masses. It is important to note at this
point that many stellar evolution calculations such as those by Pylyser \& Savonije (1988) or Alberts et
al (1996) cannot be used to set this limit, because they make use of a stellar evolution code which
cannot resolve the shell flashes described by Iben \& Tutukov.}, the 0.15
M$_{\odot}$ models do dredge up some helium, becoming mixed atmosphere 
 stars for a short while until the base
of the
convective zone retreats again. Figure~\ref{3Cool4} shows a diagram similar to Figure~\ref{3Cool1} but
for a 0.15 M$_{\odot}$ model. We see that the base of the convective zone penetrates to the
helium layer, leading to an atmospheric helium abundance of $\sim 15 \%$ in this case. Hence,
for about 1.5 Gyr, the star would exhibit the characteristics of a cool H/He atmosphere
 star, although,
with a temperature of 4000-5000 K, it would be difficult to identify it as such. For the thick
hydrogen layer models, only the lowest mass models are affected by atmospheric helium contamination.
In the sequence of models with thin ($\sim 10^{-6} \rm M_{\odot}$) hydrogen masses, dredge-up
occurs for all masses. In Figure~\ref{DAB} we show the effective temperature ranges in which this
occurs for all model masses.


Figures~\ref{Lt} and \ref{Tt} show the cooling sequences for six different models spanning the 
range of representative masses. We show curves for 0.15, 0.25 and 0.45 $\rm M_{\odot}$, with
two different hydrogen envelope masses, $3 \times 10^{-4} \rm M_{\odot}$ (thick) and $10^{-6} 
\rm M_{\odot}$ (thin). For completeness we also show the sequences for pure helium models
in Figure~\ref{Heseq}. In principle, as found by D'Antona \& Mazzitelli (1989), one can obtain
shorter cooling times for these models, but the opacities for $ T_{\rm eff} < 6000$ K are rather uncertain, and
thus we do not expect the cooling curves to be accurate at temperatures below this value. 


In order to make comparison with observations easier, we have also calculated the black-body
absolute magnitudes for these sequences, using the flux calibrations of Bessell (1979).
The most common bands used are V and I. By comparison with the detailed atmospheric calculations
of Bergeron, Saumon \& Wesemael (1995), we find that this is an excellent approximation for
$T_{\rm eff} > 4500$~K. This can be understood by comparing the Rosseland \& Planck mean opacities
shown in Figure~\ref{match} which correspond to different frequency weightings of the true monochromatic
opacity. When they are equal this indicates an approximately uniform continuum opacity and the black body
approximation is valid. For $T_{\rm eff} < 4500$~K, the CIA of H$_2$ begins to dominate and this leads
to significant deviations from the black body approximation, as can be seen in Figure~\ref{match}.
Although the Bergeron et al results are calculated for higher gravities than we require, they represent
an upper limit on the deviations from a black-body. This is because CIA is an intrinsically high
density phenomenon (it requires collisional interactions) and photospheric densities are lower in lower
gravity atmospheres. This can be demonstrated by comparing the low temperature results for
the $\log~g=7.5$ and $\log~g=8.5$ atmospheres of Bergeron et al. We will discuss the uncertainties again
in paper~II when we come to apply these models.

In Tables~\ref{Look1}, \ref{Look2} and \ref{Look3} we present the model parameters
(mass, age and gravity) as a function of the observable quantities $T_{\rm eff}$ and $M_I$.
Each entry contains three numbers, the mass in $\rm M_{\odot}$, the age in Gyr and log~$g$.
Figures~\ref{MV} and \ref{MI} show $M_V$ and $M_I$ as a function of age.



For stars with hydrogen envelopes and effective temperatures $> 7000$ K, spectroscopic determinations of effective temperature
and gravity can provide a direct measurement of the white dwarf mass (Bergeron, Wesemael \&
Fontaine 1991) , provided one
has a relationship between radius and mass. We have calculated the mass-radius relations
for both our thick and thin hydrogen layer models for the full range of helium core white dwarf
masses. We fit this by a relation 
 between gravity $ g = G M/R^2$ and effective
temperature $ T_{\rm eff}$ for a given mass, namely 

\be
 g  = F(T_{\rm eff})
 \left[ a_2 - a_3 T_{eff} \right], \label{GT} \ee
with
\be
F(T_{\rm eff}) =  1 + \frac{a_1 T_{\rm eff}}{1 + 9\, exp \left( 4 \times 10^{-3} \left( T_{\rm eff} - 5800 K
\right) \right) }, 
\ee 
and
\begin{eqnarray} 
 a_1 & = &  9.91 \times 10^{-7} \left( \frac{M}{\rm M_{\odot}} \right)^{-2.33}, \label{a1} \\
 a_2 & = &  1.69 \times 10^8 \left( \frac{M}{\rm M_{\odot}} \right)^{1.68}, \label{a2}\\
 a_3 & = &  946 \left( \frac{ M}{\rm M_{\odot}} \right)^{0.17} \label{a3}
\end{eqnarray}
for the thick hydrogen envelope and
\begin{eqnarray} 
 a_1 & = &  3.79 \times 10^{-7} \left( \frac{M}{\rm M_{\odot}} \right)^{-2.18}, \\
 a_2 & = &  1.68 \times 10^8 \left( \frac{M}{\rm M_{\odot}} \right)^{1.55}, \\
 a_3 & = &  650 
\end{eqnarray}
for the thin hydrogen envelope and $M \leq 0.4 \, \rm M_{\odot}$. Figure~\ref{gfit} compares
the fits to the proper curves for the case of a thick envelope.



We may compare the $ T_{\rm eff}=0$ limit of (\ref{GT}) with various well-known $T=0$
mass-radius relations.
 Converting the above into
mass-radius relations and extrapolating to the $T=0$ limit we get
\begin{eqnarray}
\frac{R}{\rm R_{\odot}} & = & 0.013 \left( \frac{M}{\rm M_{\odot}} \right)^{-0.32} \, \, (\rm Thick \, H),
\label{thickR}
\\
\frac{R}{\rm R_{\odot}} & = & 0.013 \left( \frac{M}{\rm M_{\odot}} \right)^{-0.28} \, \, (\rm Thin \, H).
\label{thinR}
\end{eqnarray}
This is very close to the often used mass-radius relationship of
 Paczynski (1967) for low mass degenerate dwarfs
$ R/{\rm R_{\odot}} = 0.013 \left( M/{\rm M_{\odot}} \right)^{-1/3}. $
In Figure~\ref{HS} we compare our $T=0$ curves with the Paczynski relation and that of
 Hamada and
Salpeter (1961).

\section{Conclusion}

We have presented a set of cooling sequences for low mass helium white dwarfs of different
masses and with different masses of surface hydrogen. We provide blackbody absolute magnitudes and
surface gravity - effective temperature relations as an aid to the analysis of future observations. 
In paper~II we shall apply these models to the optical 
 observations of the companions to millisecond
 pulsars in order to derive cooling ages.

The authors would like to thank Marten van Kerkwijk and Yanqin Wu for lengthy
discussions about white dwarf physics and Glenn Soberman for helping with the 
initial conditions
for our models. The generosity of messrs D. Saumon, G. Fontaine, I. Mazzitelli, 
F. Rogers and C. Iglesias in
providing their microphysical results is also appreciated. We would also like to
thank the referee, Dr. D'Antona, for some insightful comments on the atmospheric physics.

\begin{appendix}
\section{Brief Numerical Details}
\label{NumDet}
It is our intention to make both the results and the cooling code publicly accessible.
To obtain the most up-to-date version of the code email {\em hansen@cita.utoronto.ca}. In this
brief appendix we outline the version of the stellar structure equations that our program
is designed to solve.

The four stellar structure equations are
\begin{eqnarray}
\frac{{\rm d}P}{{\rm d}r} & = & - \rho \frac{G M(r)}{r^2}, \label{SS1} \\
\frac{{\rm d}M}{{\rm d}r} & = & 4 \pi r^2 \rho,   \label{SS2} \\
\frac{{\rm d}L}{{\rm d}r} & = & 4 \pi r^2 \rho \epsilon, \label{SS3} \\
\frac{{\rm d}T}{{\rm d}P} & = & \frac{T}{P} \nabla, \label{SS4}
\end{eqnarray}
where $\nabla$ in equation~(\ref{SS4}) is determined by
\be  \nabla = \frac{3}{16\pi a c}\frac{\kappa }{T^3} \frac{L(r)}{G M(r)} 
\label{radgrad}
 \ee
for a radiative atmosphere or from a mixing length calculation for a convective
atmosphere. The function $\epsilon$ contains both the nuclear energy generation terms and
the thermal and gravitational loss terms, as well as the neutrino energy losses.
 We use the transformation of variables from (see Table~\ref{Symbols} for definition of
symbols)
\begin{eqnarray}
 x & = &  r/10^{11} \, cm, \\
 \ell & = & L/L_0, \\
 \pi & = & \ln P/P_0, \\
 \theta & = & \ln T/T_0, \\
 \phi & = & \ln \rho/\rho_0, \\
 \bar{\epsilon} & = & \frac{L_0}{M_0} \epsilon, \\
 \tau & = &  t \frac{L_0 \rho_0}{M_0 P_0}, \\
 m & = &  M/M_0 = 1 - {\rm exp}(-\xi),
\end{eqnarray}
with the normalizations 
\begin{eqnarray}
 P_0 & = &  2.123 \times 10^{14} \, {\rm dyn \, cm^{-2}} \left( \frac{M_0}{\rm M_{\odot}} \right)^2, \\
 \rho_0 & = & 0.159 \, {\rm g \, cm^{-3}} \, \frac{M_0}{\rm M_{\odot}}, \\
 T_0 & = & 3.568 \times 10^{6} \, {\rm K} \left( \frac{M_0}{\rm M_{\odot}} \frac{L_0}{\rm L_{\odot}}
\right)^{1/4}, \\
\rm and && \\
 L_0 & = &  100 \, \rm L_{\odot}. 
\end{eqnarray}
The set of equations that we actually solve numerically is
\begin{eqnarray}
\frac{{\rm d}x}{{\rm d}\xi} & = & \frac{e^{\xi-\phi}}{x^2}, \\
\frac{{\rm d}\ell}{{\rm d}\xi} & = &  e^{\xi} \left( \frac{Q}{\nabla_{\rm ad}} e^{\pi - \phi} \left[
\nabla_{ad} \frac{{\rm d}\pi}{{\rm d}\tau} - \frac{{\rm d}\theta}{{\rm d}\tau} \right] - \right.
\nonumber \\
& & \left. \bar{\epsilon}_{\nu} + 
\bar{\epsilon}_{\rm nuc} \right), \\
\frac{{\rm d}\pi}{{\rm d}\xi} & = &  - \left( 1 - e^{-\xi} \right) \frac{e^{\xi - \pi}}{x^4}, \\
\frac{{\rm d}\theta}{{\rm d}\xi} & = & - \left( 1 - e^{-\xi} \right) \frac{e^{\xi - \pi}}{x^4} \nabla, 
\end{eqnarray}
(Q =  - $ \left( \frac{\partial \ln \rho}{\partial \ln T} \right)_P$, $\rm \nabla_{\rm ad}$ = 
$\rm \left( \frac{\partial \ln T}{\partial \ln P} \right)_S$).
This system of equations is solved on a grid equally spaced in $\xi$. The external
boundary condition is determined by integrating inwards using an adaptive
stepsize algorithm starting with a 
 grey atmosphere solution 
to determine $\pi$ and $\theta$ at the outer grid point as a function of $\ell$ and $x$. The
fitting procedure used was taken from the triangle method of Kippenhahn et al. (1967).

\end{appendix}

\label{lastpage}
\clearpage

\begin{table*}
\begin{minipage}{200mm}
\caption{Masses of helium Core white dwarfs ($\rm M_{\odot}$) \label{Look1}}
\begin{centering}
\begin{tabular}{lcccccccccccccccc}
T$_{\rm eff}$ & \multicolumn{16}{c}{$ M_I$} \\ 
(K)& 10.0 & 10.5 & 11.0 & 11.5 & 12.0 & 12.5 & 12.75 &
13.0 & 13.25 & 13.5 & 13.75 & 14.0 & 14.25 & 14.50 &
14.75 & 15.0  \\
\hline
12000 & 0.27 & 0.36 &      &      &      &      &      &      &      &      &      &      &      &      &      &     
 \\
11500 & 0.25 & 0.33 & 0.45 &      &      &      &      &      &      &      &      &      &      &      &      &     
 \\
11000 & 0.24 & 0.30 & 0.43 &      &      &      &      &      &      &      &      &      &      &      &      &     
 \\
10500 & 0.22 & 0.28 & 0.40 &      &      &      &      &      &      &      &      &      &      &      &      &     
 \\
10000 & 0.21 & 0.26 & 0.36 &      &      &      &      &      &      &      &      &      &      &      &      &     
 \\
 9500 & 0.19 & 0.24 & 0.32 & 0.45 &      &      &      &      &      &      &      &      &      &      &      &     
 \\
 9000 & 0.18 & 0.22 & 0.29 & 0.42 &      &      &      &      &      &      &      &      &      &      &      &     
 \\
 8500 & 0.17 & 0.20 & 0.26 & 0.38 &      &      &      &      &      &      &      &      &      &      &      &     
 \\
 8000 & 0.16 & 0.18 & 0.23 & 0.32 &      &      &      &      &      &      &      &      &      &      &      &     
 \\
 7500 &      & 0.16 & 0.20 & 0.27 & 0.42 &      &      &      &      &      &      &      &      &      &      &     
 \\
 7000 &      &      & 0.18 & 0.23 & 0.34 &      &      &      &      &      &      &      &      &      &      &     
 \\
 6500 &      &      & 0.15 & 0.19 & 0.27 & 0.43 &      &      &      &      &      &      &      &      &      &     
 \\
 6000 &      &      &      & 0.16 & 0.21 & 0.33 & 0.42 &      &      &      &      &      &      &      &      &     
 \\
 5800 &      &      &      &      & 0.19 & 0.29 & 0.37 & 0.44 &      &      &      &      &      &      &      &     
 \\
 5600 &      &      &      &      & 0.17 & 0.25 & 0.32 & 0.41 &      &      &      &      &      &      &      &     
 \\
 5400 &      &      &      &      & 0.15 & 0.21 & 0.27 & 0.35 & 0.44 &      &      &      &      &      &      &     
 \\
 5200 &      &      &      &      &      & 0.18 & 0.22 & 0.29 & 0.39 &      &      &      &      &      &      &     
 \\
 5000 &      &      &      &      &      &      & 0.19 & 0.24 & 0.32 & 0.42 &      &      &      &      &      &     
 \\
 4800 &      &      &      &      &      &      & 0.15 & 0.20 & 0.26 & 0.34 & 0.44 &      &      &      &      &     
 \\
 4600 &      &      &      &      &      &      &      & 0.16 & 0.20 & 0.27 & 0.37 &      &      &      &      &     
 \\
 4400 &      &      &      &      &      &      &      &      & 0.16 & 0.21 & 0.28 & 0.39 &      &      &      &     
 \\
 4200 &      &      &      &      &      &      &      &      &      & 0.16 & 0.22 & 0.29 & 0.40 &      &      &     
 \\
 4000 &      &      &      &      &      &      &      &      &      &      & 0.16 & 0.22 & 0.29 & 0.40 &      &     
 \\
 3800 &      &      &      &      &      &      &      &      &      &      &      & 0.16 & 0.21 & 0.29 & 0.40 &     
 \\
 3600 &      &      &      &      &      &      &      &      &      &      &      &      &      & 0.20 & 0.27 & 0.37
 \\
 3400 &      &      &      &      &      &      &      &      &      &      &      &      &      &      & 0.18 & 0.24
 \\
\hline
\end{tabular}
\end{centering}

The values in this table are for a thick hydrogen envelope ($3 \times 10^{-4} \rm M_{\odot}$).
\end{minipage}
\end{table*}

\begin{table*}
\begin{minipage}{200cm}
\caption{Ages of helium core white dwarfs (Gyr) \label{Look2}}
\begin{centering}
\begin{tabular}{lcccccccccccccccc}
$T_{\rm eff}$ & \multicolumn{16}{c}{$M_I$} \\
(K) & 10.0 & 10.5 & 11.0 & 11.5 & 12.0 & 12.5 &
12.75 & 13.0 & 13.25 & 13.5 & 13.75 & 14.0 & 14.25 &
14.5 & 14.75 & 15.0  \\
\hline
12000 &  0.2 &  0.4 &      &      &      &      &      &      &      &      &      &      &      &      &      &     
 \\
11500 &  0.2 &  0.4 &  0.7 &      &      &      &      &      &      &      &      &      &      &      &      &     
 \\
11000 &  0.2 &  0.4 &  0.8 &      &      &      &      &      &      &      &      &      &      &      &      &     
 \\
10500 &  0.2 &  0.4 &  0.8 &      &      &      &      &      &      &      &      &      &      &      &      &     
 \\
10000 &  0.2 &  0.4 &  0.7 &      &      &      &      &      &      &      &      &      &      &      &      &     
 \\
 9500 &  0.1 &  0.4 &  0.7 &  1.1 &      &      &      &      &      &      &      &      &      &      &      &     
 \\
 9000 &  0.1 &  0.3 &  0.6 &  1.2 &      &      &      &      &      &      &      &      &      &      &      &     
 \\
 8500 &  0.1 &  0.3 &  0.6 &  1.1 &      &      &      &      &      &      &      &      &      &      &      &     
 \\
 8000 &  0.1 &  0.3 &  0.6 &  1.1 &      &      &      &      &      &      &      &      &      &      &      &     
 \\
 7500 &      &  0.3 &  0.5 &  1.0 &  1.8 &      &      &      &      &      &      &      &      &      &      &     
 \\
 7000 &      &      &  0.5 &  0.9 &  1.6 &      &      &      &      &      &      &      &      &      &      &     
 \\
 6500 &      &      &  0.4 &  0.8 &  1.5 &  2.6 &      &      &      &      &      &      &      &      &      &     
 \\
 6000 &      &      &      &  0.7 &  1.3 &  2.3 &  3.1 &      &      &      &      &      &      &      &      &     
 \\
 5800 &      &      &      &      &  1.2 &  2.1 &  2.9 &  3.9 &      &      &      &      &      &      &      &     
 \\
 5600 &      &      &      &      &  1.1 &  2.0 &  2.6 &  3.7 &      &      &      &      &      &      &      &     
 \\
 5400 &      &      &      &      &  1.0 &  1.8 &  2.4 &  3.4 &  4.9 &      &      &      &      &      &      &     
 \\
 5200 &      &      &      &      &      &  1.6 &  2.2 &  3.1 &  4.6 &      &      &      &      &      &      &     
 \\
 5000 &      &      &      &      &      &      &  2.0 &  2.8 &  4.0 &  5.9 &      &      &      &      &      &     
 \\
 4800 &      &      &      &      &      &      &  1.8 &  2.5 &  3.5 &  5.1 &  7.1 &      &      &      &      &     
 \\
 4600 &      &      &      &      &      &      &      &  2.2 &  3.1 &  4.4 &  6.4 &      &      &      &      &     
 \\
 4400 &      &      &      &      &      &      &      &      &  2.7 &  3.7 &  5.3 &  7.7 &      &      &      &     
 \\
 4200 &      &      &      &      &      &      &      &      &      &  3.1 &  4.3 &  6.1 &  8.9 &      &      &     
 \\
 4000 &      &      &      &      &      &      &      &      &      &      &  3.4 &  4.8 &  6.9 & 10.0 &      &     
 \\
 3800 &      &      &      &      &      &      &      &      &      &      &      &  3.7 &  5.3 &  7.6 & 10.9 &     
 \\
 3600 &      &      &      &      &      &      &      &      &      &      &      &      &      &  5.5 &  7.9 & 11.3
 \\
 3400 &      &      &      &      &      &      &      &      &      &      &      &      &      &      &  5.5 &  7.9
 \\
\hline
\end{tabular}
\end{centering}

The values in this table are for a thick hydrogen envelope ($3 \times 10^{-4} \rm M_{\odot}$).
\end{minipage}
\end{table*}

\begin{table*}
\begin{minipage}{200cm}
\caption{Gravities of helium core white dwarfs (log~$g$) \label{Look3}}
\begin{centering}
\begin{tabular}{lcccccccccccccccc}
$T_{\rm eff}$ & \multicolumn{16}{c}{$M_I$} \\
(K) & 10.0 & 10.5 & 11.0 & 11.5 & 12.0 & 12.5 &
12.75 & 13.0 & 13.25 & 13.5 & 13.75 & 14.0 & 14.25 &
14.5 & 14.75 & 15.0  \\
\hline
12000 & 7.01 & 7.33 &      &      &      &      &      &      &      &      &      &      &      &      &      &     
 \\
11500 & 6.95 & 7.27 & 7.60 &      &      &      &      &      &      &      &      &      &      &      &      &     
 \\
11000 & 6.88 & 7.19 & 7.54 &      &      &      &      &      &      &      &      &      &      &      &      &     
 \\
10500 & 6.81 & 7.12 & 7.47 &      &      &      &      &      &      &      &      &      &      &      &      &     
 \\
10000 & 6.74 & 7.04 & 7.38 &      &      &      &      &      &      &      &      &      &      &      &      &     
 \\
 9500 & 6.66 & 6.95 & 7.28 & 7.62 &      &      &      &      &      &      &      &      &      &      &      &     
 \\
 9000 & 6.57 & 6.85 & 7.18 & 7.55 &      &      &      &      &      &      &      &      &      &      &      &     
 \\
 8500 & 6.50 & 6.75 & 7.07 & 7.44 &      &      &      &      &      &      &      &      &      &      &      &     
 \\
 8000 & 6.39 & 6.65 & 6.95 & 7.30 &      &      &      &      &      &      &      &      &      &      &      &     
 \\
 7500 &      & 6.53 & 6.82 & 7.16 & 7.54 &      &      &      &      &      &      &      &      &      &      &     
 \\
 7000 &      &      & 6.68 & 7.00 & 7.37 &      &      &      &      &      &      &      &      &      &      &     
 \\
 6500 &      &      & 6.53 & 6.83 & 7.18 & 7.57 &      &      &      &      &      &      &      &      &      &     
 \\
 6000 &      &      &      & 6.65 & 6.97 & 7.35 & 7.56 &      &      &      &      &      &      &      &      &     
 \\
 5800 &      &      &      &      & 6.87 & 7.25 & 7.45 & 7.64 &      &      &      &      &      &      &      &     
 \\
 5600 &      &      &      &      & 6.78 & 7.13 & 7.34 & 7.55 &      &      &      &      &      &      &      &     
 \\
 5400 &      &      &      &      & 6.67 & 7.01 & 7.22 & 7.43 & 7.62 &      &      &      &      &      &      &     
 \\
 5200 &      &      &      &      &      & 6.88 & 7.07 & 7.29 & 7.51 &      &      &      &      &      &      &     
 \\
 5000 &      &      &      &      &      &      & 6.93 & 7.14 & 7.36 & 7.58 &      &      &      &      &      &     
 \\
 4800 &      &      &      &      &      &      & 6.77 & 6.98 & 7.20 & 7.42 & 7.63 &      &      &      &      &     
 \\
 4600 &      &      &      &      &      &      &      & 6.81 & 7.03 & 7.25 & 7.48 &      &      &      &      &     
 \\
 4400 &      &      &      &      &      &      &      &      & 6.84 & 7.07 & 7.29 & 7.52 &      &      &      &     
 \\
 4200 &      &      &      &      &      &      &      &      &      & 6.86 & 7.08 & 7.31 & 7.55 &      &      &     
 \\
 4000 &      &      &      &      &      &      &      &      &      &      & 6.86 & 7.09 & 7.32 & 7.56 &      &     
 \\
 3800 &      &      &      &      &      &      &      &      &      &      &      & 6.84 & 7.07 & 7.30 & 7.55 &     
 \\
 3600 &      &      &      &      &      &      &      &      &      &      &      &      &      & 7.02 & 7.26 & 7.50
 \\
 3400 &      &      &      &      &      &      &      &      &      &      &      &      &      &      & 6.94 & 7.18
 \\
\hline
\end{tabular}
\end{centering}

The values in this table are for a thick hydrogen envelope ($3 \times 10^{-4} \rm M_{\odot}$).
\end{minipage}
\end{table*}

\begin{table*}
\begin{minipage}{140mm}
\caption{Symbols used in Appendix \label{Symbols}}
\begin{centering}
\begin{tabular}{lcc}
Physical quantity & Symbol & Transformed \\
\hline
Pressure & $P$ & $\pi$ \\
Temperature & $T$ & $\theta$ \\
Luminosity & $L$ & $\ell$ \\
Radius & $r$ & $x$ \\
Mass & $m$ & $\xi$ \\
Time & $t$ & $\tau$ \\
Density & $\rho$ & $\phi$ \\
 \hline
\end{tabular}
\end{centering}
\end{minipage}
\end{table*}

\clearpage
\begin{figure}
\caption{{\bf The Hydrogen Phase Diagram:}
The thin solid lines are contours of constant log $\kappa$, where $\kappa$ is the
Rosseland mean opacity
in $\rm cm^2 \, g^{-1}$. The contours have values 0, 2 and 4. The dotted lines are also contours of
constant
log $\kappa$ but with values -2, -4, etc., down to -12 (decreasing monotonically in all directions
from the opacity peak near 30000~K and $10^{-3} \rm g.cm^{-3}$). The heavy dashed lines delineate regions where
different tables have been used to calculate the opacity. The Itoh opacities are valid for
$T < 0.1 \,  T_{\rm F}$ (where $ T_{\rm F}$ is the Fermi temperature), $y <0.1$ ($y$ measures the importance of
the
wave nature of the ions) and $\Gamma < 171$. There is also a
lower bound on the density $\rho > 100 \, \rm g \, cm^{-3}$. Outside of this region we use the 
conductivities
of Hubbard and Lampe. The radiative opacities in the region T $\sim 10^4 - 10^5$ K and $\rho \sim
10^{-4} - 1\, \rm g \, cm^{-3}$ were obtained by extrapolating the OPAL opacities to higher densities. 
This
extrapolation is not important because the atmosphere is convective at these temperatures and
densities. The box in the lower left-hand corner is the region covered by our opacity calculations.
We can see the opacity minimum near 3000 K due to the change in the dominant opacity mechanism
from $\rm H^-$ absorption to $\rm H_2$ CIA. The strange behaviour in the upper left-hand corner is due 
to
the extrapolation of the conductive opacities outside their range of validity. This is unimportant
as no model we consider will approach this region.}
\label{hydrogen}
\end{figure}
\begin{figure}
\caption{{\bf Matching Opacities:}
 The solid points are Rosseland mean opacities calculated using our code. The open
squares are the results of the OPAL calculation. The open stars are again our calculation but
showing Planck mean opacities. The vertical dotted lines delineate the region 6000-7000 K which
is where the two calculations overlap. Once again, the minimum in the opacity near 3000 K is
due to the change in the dominant opacity contributor, from H$_2$ CIA at lower temperatures to
H$^-$ absorption at higher temperatures. We see that below 5000 K there is a significant
discrepancy between the Planck and Rosseland mean opacities.}
\label{match}
\end{figure}
\begin{figure}
\caption{{\bf Code Comparisons: C/O models:}
The open squares denote the models we compare against. The filled squares are our
own models. The arrows denote the point at which the core of the model begins to crystallise,
although some of the difference at low luminosities is due to the updated opacities.
The left panel describes a 0.564 $\rm M_{\odot}$ oxygen core surrounded by a helium envelope of
$2.5 \times 10^{-3} \rm M_{\odot}$ and a hydrogen envelope of $3 \times 10^{-4} \rm M_{\odot}$. The
metallicity is taken to be $Z=0$. The centre panel describes a 0.546 $\rm M_{\odot}$ carbon core with
a helium envelope of $0.022 \, \rm M_{\odot}$ and hydrogen envelope $10^{-4} \rm M_{\odot}$. The
metallicity
is $Z=0.02$. The rightmost panel is a 0.6 $\rm M_{\odot}$ star, with a carbon core, mass
fraction $10^{-2}$ of helium and $10^{-4}$ of hydrogen. The metallicity is
$Z=0$. We had to adjust our conductive opacities to reproduce the above
results. When prior authors used Hubbard \& Lampe opacities in regions where we used Itoh opacities,
we divided our opacities by a factor of 2 to compensate.}
\label{compare}
\end{figure}

\begin{figure}
\caption{ {\bf Code Comparisons: He model:}
 The solid line shows our model with no corrections for different starting points.
The dashed line corresponds to the same model, but with the age incremented by $10^8$ years,
to compensate for the time spent in prior evolutionary stages. The open squares are the
results of Iben \& Tutukov (1986). The agreement is excellent until $t \sim$ 2 Gyr, by which
point Iben \& Tutukov caution that their opacities are uncertain.}
\label{it}
\end{figure}

\begin{figure}
\caption{{\bf Cooling of a 0.3 $\rm M_{\odot}$ Model:}
 We show here the cooling of a 0.3 $\rm M_{\odot}$ star with a hydrogen
envelope of $3 \times 10^{-4} \rm M_{\odot}$. The solid line is the electromagnetic bolometric
luminosity $L_{\gamma}$, the dotted lines indicate the neutrino and nuclear luminosities
respectively (the nuclear contribution is included in the bolometric luminosity) and the
short and long dashed lines indicate the gravity (in units of $ \rm 10^7 cm \, s^{-2}$) and
the central degeneracy ($\eta_{\rm c} = E_{\rm F}/kT$) respectively. The effective temperatures
corresponding to the various ages for this model are shown on the top axis, ranging from
20000~K at the left to 3000~K at the right.}
\label{3Cool3}
\end{figure}

\begin{figure}
\caption{{\bf Evolution in the Phase Diagram:}
This $T-P$ phase diagram shows three representative atmosphere profiles (labelled at the top by
their age in Gyr)
for the evolution of the same model shown in \protect{\ref{3Cool3}}. The
heavy solid lines indicate the helium parts of the star and the thin solid
lines indicate the hydrogen part. The dotted lines delineate the
regions of 50-50 division between HI-HII and $\rm H_2$-HI respectively. The
dashed line indicates the boundary of the convective region for this model
(the other pair of dashed lines in the upper left-hand corner indicates crystallization
boundary of helium). The thick solid line at the lower left indicates the location of
 the photosphere for
this cooling sequence. The labelled dashes on each of the three curves
 indicate the points at which the degeneracy parameter $\eta =  E_{\rm F}/kT$ has
that particular value.}
\label{3Cool2}
\end{figure}

\begin{figure}
\caption{{\bf 0.3 $\rm M_{\odot}$ Convective Zone:}
 Here we show the mass in the convective zone as a function of age (or effective
temperature). The shaded region is the convective zone, while the heavy solid line indicates the
location of the photosphere. The dashed line indicates the hydrogen-helium interface in this set of
models. We note the appearance of a radiative buffer zone at late times associated with the
transition from $\rm H^-$ opacity to $\rm H_2$ opacity.}
\label{3Cool1}
\end{figure}

\begin{figure}
\caption{{\bf 0.15 $\rm M_{\odot}$ Convective Zone:}
Once again the shaded region is the convective zone, and the heavy solid line
denotes the position of the photosphere. We note that, for $T_{\rm eff} \sim 4000-5000$ K,
the atmosphere will be contaminated with helium.}
\label{3Cool4}
\end{figure}
                                                                   
\begin{figure}
\caption{{\bf Atmospheric Helium Contamination for Thin Hydrogen Envelopes:}
 The shaded regions indicate those models in which the convective zone extends into
the helium layer and thus causes atmospheric helium contamination. We consider two representative
cases. The area marked as H/He is characterised by a mass fraction of helium, $X_{\rm He}$ $> 0.1$.
The area marked as He/H is characterised by $ X_{\rm He} > 0.8$. This cutoff value may seem rather
high, but we note that the dredge-up occurs for temperatures at which helium is neutral and thus
makes little contribution to the opacity. The horizontal dotted lines indicate the mass limits
of the models we calculated, so that the extent of the convective regions outside these bounds
is unknown. These models are for a hydrogen envelope of mass $ M_{\rm H} = 10^{-6} \, \rm M_{\odot}$.}
\label{DAB}
\end{figure}

\begin{figure}
\caption{{\bf Hydrogen Cooling Sequences 1: Luminosity Evolution:}
The solid lines indicate model white dwarf
cooling sequences with
a hydrogen envelope of $3 \times 10^{-4} \rm M_{\odot}$ for each of three representative
total masses. The dashed lines are the equivalent sequences with a smaller hydrogen
envelope of $10^{-6} \rm M_{\odot}$. The difference in luminosities at earlier times
is a result of the thicker hydrogen layer leading to a larger stellar radius (the
effective temperatures are closer - see Figure~\protect{\ref{Tt}}).}
\label{Lt}
\end{figure}

\begin{figure}
\caption{{\bf Hydrogen Cooling Sequences 2: Temperature Evolution:}
 Here we show the effective temperature for the same sequences as in
Figure~\protect{\ref{Lt}}. Note the large variation in temperature with envelope mass for the
most massive models. This is the effect of the contribution of residual hydrogen burning at the
base of the thicker hydrogen envelope.}
\label{Tt}
\end{figure}

\begin{figure}
\caption{{\bf Helium Cooling Sequences:}
 We show here pure helium models for the same masses as before. The evolution at
effective temperatures below 6000 K is uncertain because of the inaccuracy of the
photospheric opacities for neutral helium at these temperatures.}
\label{Heseq}
\end{figure}
                           
\begin{figure}
\caption{{\bf V Band Cooling Sequences:}
We show here the absolute V magnitude determined from our cooling sequences. The
curves are for 0.15, 0.25, 0.35 and 0.45 $\rm M_{\odot}$.}
\label{MV}
\end{figure}

\begin{figure}
\caption{{\bf I Band Cooling Sequences:}
As for Figure~\protect{\ref{MV}}, but for absolute I magnitude.}
\label{MI}
\end{figure}

\begin{figure}
\caption{{\bf The Gravity-Effective Temperature Relation:}
The solid lines represent the true $g- T_{\rm eff}$ curves, and the
dashed lines are the fits given by equations (\protect{\ref{GT}})-(\protect{\ref{a3}}).
\label{gfit}}
\end{figure}

\begin{figure}
\caption{{ \bf The $T=0$ Mass-Radius Relation:}
The dotted line is the gravity as determined from the Paczynski (1967) mass-radius
relation. The dashed line was obtained using the Hamada and Salpeter (1961) pure helium
mass-radius relation. The filled circles are for the thick H envelope models
(equation~\protect{\ref{thickR}}) and the open circles, for the thin H envelope models
(equation~\protect{\ref{thinR}}).
\label{HS}}
\end{figure}

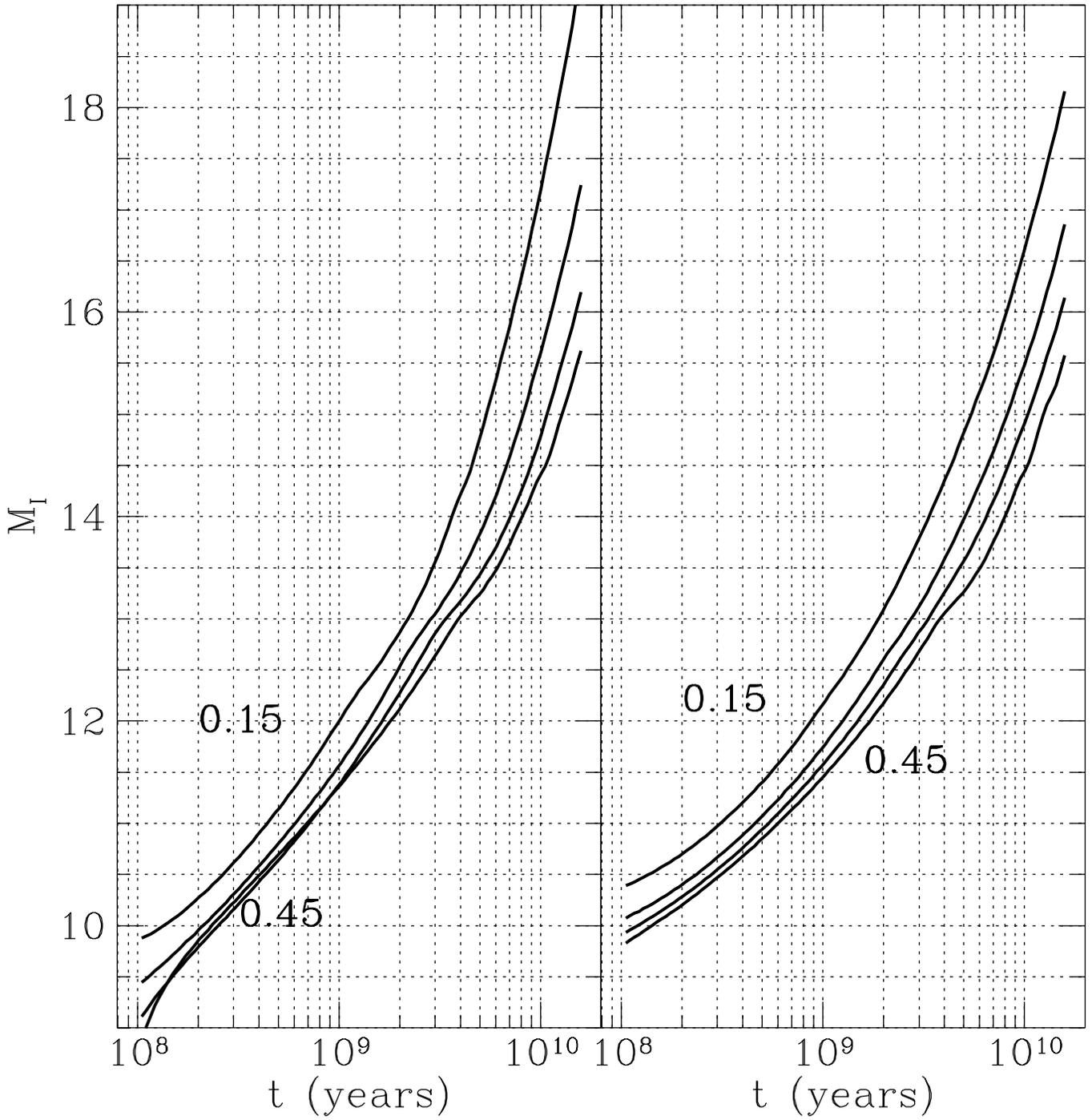
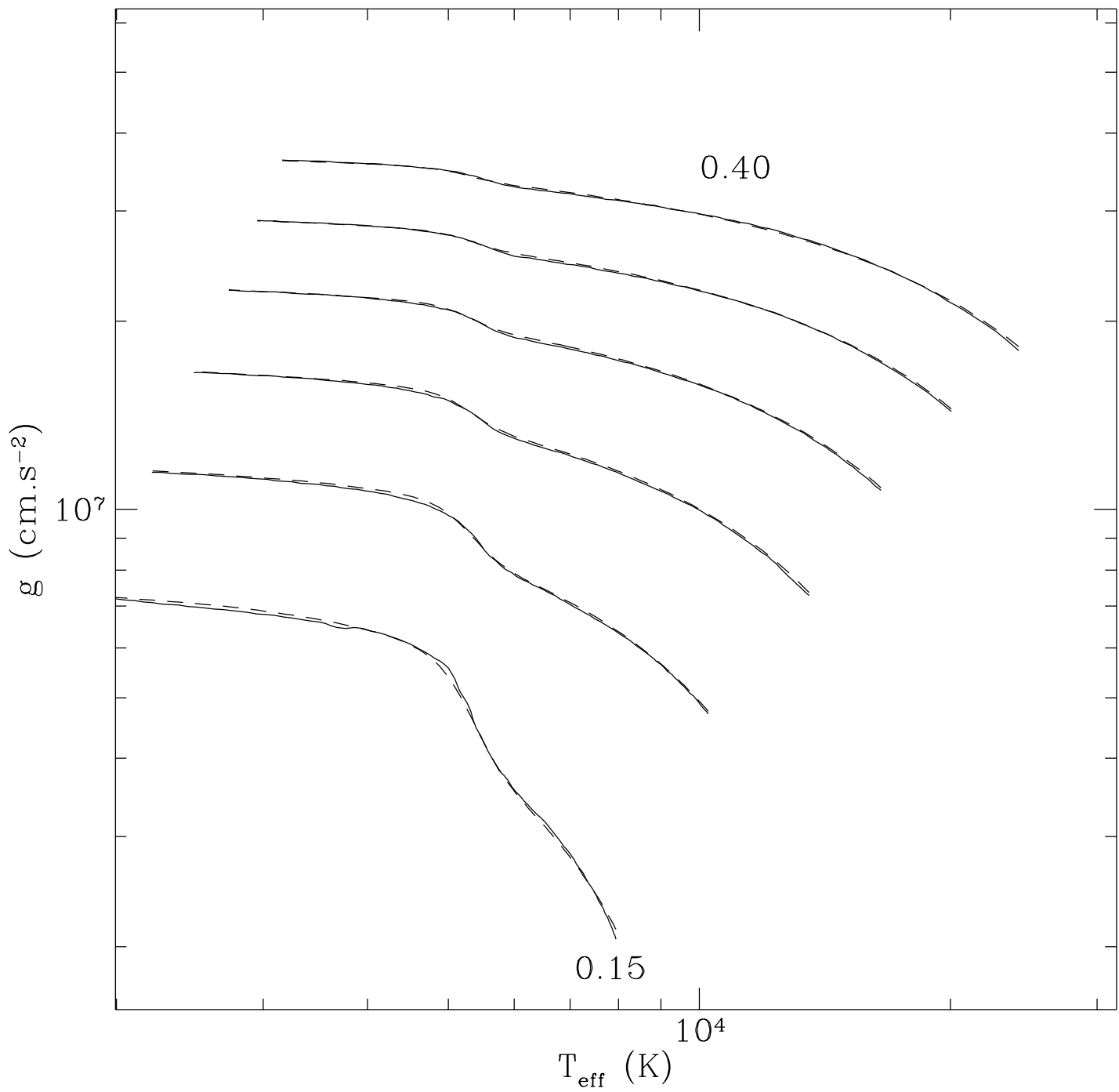
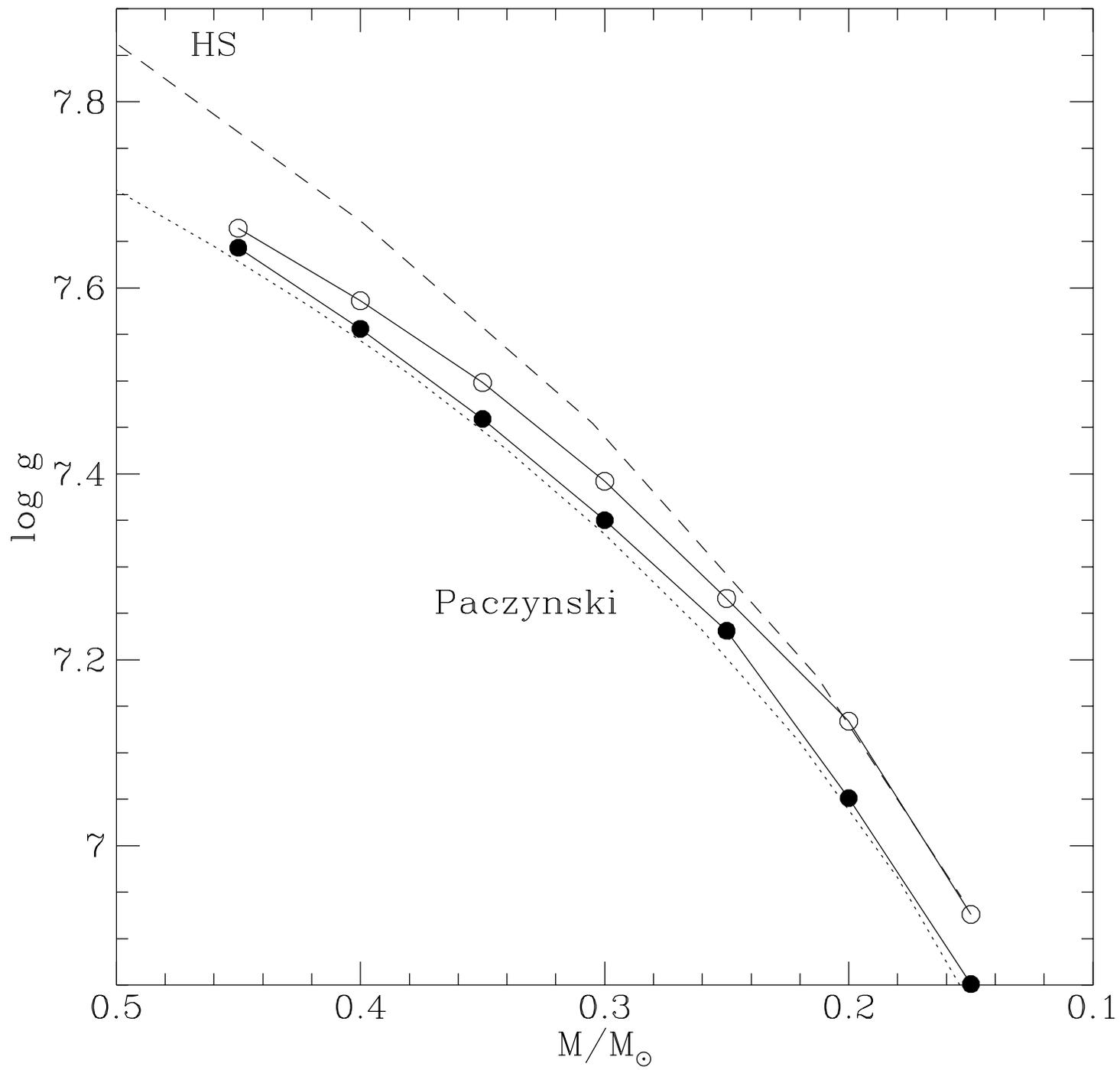
\end{document}